\documentclass{article}
\usepackage{amssymb}
\usepackage{amsfonts}
\begin{document}

\begin{center}
{\Large Contact superconformal algebras}\\
{\Large and their representations}
\end{center}
\vskip 0.1in
\begin{center}
{\large Elena Poletaeva}
\end{center}
\vskip 0.1in

{\bf 1.-}
A {\it superconformal algebra}  is a complex
${\mathbb Z}$-graded Lie superalgebra $\mathcal G = \oplus_i\mathcal G_i$, such that

\noindent
1) $\mathcal G$ is simple,

\noindent  
2) $\mathcal G$ contains the centerless Virasoro algebra 
i.e. the Lie algebra with the basis $L_n$ ($n\in\mathbb Z$) and commutation relations
$[L_m, L_n] = (m - n)L_{m+n}$ as a subalgebra,

\noindent
3) $\mathcal G_i = \lbrace x\in\mathcal G \hbox{ }|\hbox{ }
 [L_0, x] = ix\rbrace$ and dim $\mathcal G_i < C$,
where $C$ is a constant independent of $i$ [1, 2].

To describe superconformal algebras, consider
the Grassmann algebra $\Lambda(N)$ in $N$ variables
$\theta_1, \ldots, \theta_N$. Let
$\Lambda(1, N) =\mathbb C [t, t^{-1}]\otimes \Lambda (N)$ be an associative
superalgebra with natural multiplication and with the following parity 
of generators: $p(t) = \bar{0}$, $p(\theta_i) = \bar{1}$
for $i = 1, \ldots, N$. Let
$W(N)$ be the Lie superalgebra of all derivations of
$\Lambda(1, N)$.
Every $D\in W(N)$ is represented by a differential operator,
\begin{equation}
D = f\partial_t + \sum_{i=1}^N f_i \partial_i, \hbox{ where }
f, f_i \in  \Lambda(1, N).
\end{equation}
The main examples  of superconformal algebras are the following three series:
the series $W(N)$ ($N\geq 0$), the series $S'(N, \alpha)$ ($N\geq 2$)
of one-parameter families
of deformations of the divergence-free subalgebra of $W(N)$,
and the series of {\it contact superalgebras} $K'(N)$ ($N\geq 0$)([1, 2]).
By definition,
\begin{equation}
K(N) = \lbrace D \in W(N)\mid D\Omega  = f\Omega \hbox{ for some }
f\in \Lambda(1, N)\rbrace, 
\end{equation}
where
$\Omega = dt - \sum_{i=1}^N \theta_id\theta_i$
is a differential 1-form, which is called a {\it contact form} [1, 2, 3].
There is one-to one correspondence between the differential operators
$D\in K(N)$ and the functions $f \in \Lambda(1, N)$.
The correspondence $f \leftrightarrow D_f$ is given by
\begin{equation}
D_f = \Delta(f)\partial_t + \partial_t(f)\sum_{i=1}^N \theta_i\partial_i
+ (-1)^{p(f)}\sum_{i=1}^N\partial_i(f)\partial_i,
\end{equation}
where $\Delta(f) = 2f - \sum_{i=1}^N\theta_i\partial_i(f)$.
The Lie bracket in $K(N)$ is identified with the contact bracket in 
$\Lambda(1, N)$:
\begin{equation}
\lbrace f, g\rbrace = \Delta(f)\partial_t(g) - \partial_t(f)\Delta(g)
+ (-1)^{p(f)}\sum_{i=1}^N\partial_i(f)\partial_i(g), 
\end{equation}
so that $[D_f, D_g] = D_{\lbrace f, g \rbrace}$.
The superalgebras $K(N)$ are known to the physicists as
the $SO(N)$ {\it superconformal algebras} [4]. They are
simple, except when $N = 4$. If $N = 4$,
then the derived superalgebra $K'(4) = [K(4), K(4)]$ is a simple ideal in
$K(4)$ of codimension one defined from the exact sequence
\begin{equation}
0\rightarrow K'(4)\rightarrow K(4)\rightarrow \mathbb C 
D_{t^{-1} \theta_1\theta_2\theta_3\theta_4}\rightarrow 0.
\end{equation}
There exists no nontrivial 2-cocycles on $K'(N)$ if $N > 4$.
If $N \leq 3$, then there exists, up to equivalence,
one nontrivial 2-cocycle [1]. Let $\hat{K}'(N)$ be the corresponding central 
extension of ${K}'(N)$. 
$\hat{K}'(1)$ is isomorphic to the  {\it Neveu-Schwarz algebra},
and $\hat{K}'(2)$ is isomorphic to the so-called 
$N = 2$  {\it superconformal algebra} (see references in [8]).
The superalgebra $K'(4)$ has 3 independent central extensions [1].
Note that $K'(4)$ is spanned by 16 field; it is the largest 
among superconformal algebras admitting central extensions and, 
therefore is one of the most interesting ones.

The superalgebra $K'(N)$, where $N \geq 0$,
has a two-parameter family of representations in the superspace 
spanned by $2^N$ fields. In fact, $K'(N)$ acts in a natural way
on the superspace of ``densities'' of the form
$t^{\alpha}g\Omega^{\beta}$, where $g\in  \Lambda(1, N)$, and
$\alpha$ and $\beta$ are fixed
complex numbers [1]:
\begin{equation}
D_f(t^{\alpha}g\Omega^{\beta}) =
(D_f(t^{\alpha}g) + (-1)^{p(f)p(g)}2\beta t^{\alpha}g\partial_t(f))
\Omega^{\beta}.
\end{equation}
The superalgebra $K'(4)$ has in addition a one-parameter family of spinor-like tiny
irreducible representations realized on just 4 fields instead of the
usual 16. The construction of this representation is based on the
embedding of a nontrivial central extension 
of  ${K}'(4)$ into the Lie superalgebra of pseudodifferential symbols on
the supercircle   $S^{1\mid 2}$ [8]. Note that there is no analogous
representation of $K'(2N)$ realized on $2^N$ fields for $N \geq 3$.

{\bf 2.-}  
The {\it Poisson algebra} $P$ {\it of pseudodifferential symbols
on the circle} is formed by the formal series
$A(t, \xi) = \sum_{-\infty}^na_i(t) {\xi}^i,$
where $a_i(t)\in \mathbb C [t, t^{-1}]$, and the variable $\xi$ corresponds
to $\partial_t$. 
The Poisson bracket is defined as follows:
\begin{equation}
\lbrace A(t, \xi), B(t, \xi) \rbrace = 
\partial_{\xi}A(t, \xi)\partial_{t}B(t, \xi) -
\partial_tA(t, \xi)\partial_{\xi}B(t, \xi).
\end{equation}
The Poisson algebra $P$ has a well-known deformation $P_h$,
where $h\in \rbrack 0, 1]$.
The associative multiplication in the vector space $P$
is determined as follows:
\begin{equation}
A(t, \xi)\circ_h B(t, \xi) = 
\sum_{n\geq 0}\frac {h^n} {n!}\partial^n_{\xi}A(t, \xi)
\partial^n_{t}B(t, \xi). 
\end{equation}
The Lie algebra structure on $P_h$ is given by
$[A, B]_h =  A\circ_h B - B\circ_h A$, so that the family
$P_h$ contracts to $P$ [5]. $P_{h=1}$ is called 
the  {\it Lie algebra of pseudodifferential symbols
on the circle}.

Let $\Theta(N)$ be
the Grassman algebra with generators
$\theta_1,\ldots, \theta_N, \bar{\theta}_1, \ldots, \bar{\theta}_N$,
where $\bar{\theta}_i = \partial_i$ for $i = 1, \ldots, N$.
The {\it Poisson superalgebra} $P(N)$ {\it of 
pseudodifferential symbols on the supercircle}   $S^{1\mid N}$ 
has the underlying vector space
$P\otimes \Theta (N)$.
The Poisson bracket is defined as follows:
\begin{equation}
\lbrace A, B \rbrace = 
\partial_{\xi}A\partial_{t}B -
\partial_tA\partial_{\xi}B - (-1)^{p(A)}
(\sum_{i=1}^N\partial_{\theta_i}A\partial_ {\bar{\theta}_i}B
+ \partial_ {\bar{\theta}_i}A\partial_{\theta_i}B).
\end{equation}
Let $\Theta_h(N)$ be an associative superalgebra
with generators
$\theta_1,\ldots, \theta_N, \partial_1,\ldots, \partial_N$
and relations: 
$\theta_i\theta_j = - \theta_j\theta_i,
\partial_i\partial_j = - \partial_j\partial_i,
\partial_i\theta_j = h\delta_{i,j}  - \theta_j\partial_i.$
Let
$P_h(N) = P_h \otimes \Theta_h(N)$
be an associative superalgebra with the product
\begin{equation}
(A \otimes X) (B \otimes  Y) = \frac {1}{h}(A\circ_h B)\otimes (X Y),
\end{equation}
where $A, B \in P_h$, and $X, Y \in \Theta_h(N)$.
Correspondingly, the Lie bracket in $P_h(N)$ is
$[A, B]_h = AB - (-1)^{p(A)p(B)}BA$,
and $\hbox{lim}_{h\rightarrow 0} [A, B]_h = \lbrace A, B \rbrace$
for $A, B \in P_h(N)$.
There exist natural embeddings:
$W(N)\subset P(N)$ and $W(N)\subset P_h(N)$.
$P_{h=1}(N)$ is the {\it Lie superalgebra of 
pseudodifferential symbols on}   $S^{1\mid N}$.

The Lie algebra $P_{h=1}$
has two independent central extensions [5].
Analogously, there exist, up to equivalence, two 
nontrivial 2-cocycles, $c_{\xi}$ and $c_t$,
on $P_{h=1}(N)$ [8].
Let $x = \xi, t$ and $\hat{x} = t, \xi$, respectively, and
let $\log  x$ be the derivation of the ring $P_{h=1}$ defined by
\begin{equation}
[\log  x,A(t, \xi)] = \sum_{k\geq 1}
 \frac {(-1)^{k+1}}{k}\partial^k_{\hat{x}}A(t, \xi)x^{-k},
\end{equation}
(see [5]). Then
$c_x(A \otimes X, B \otimes  Y) = 
\hbox { the coefficient of } t^{-1}\xi^{-1}\theta_1\ldots\theta_N
\partial_1\ldots\partial_N$ in the expression for
$([\log x,A]\circ_{h=1} B)\otimes (X Y)$, where 
$A, B \in P_{h=1}$, and $X, Y \in \Theta_{h=1}(N)$.

There exists an embedding for $N \geq 0$:
\begin{equation}
K(2N) \subset P(N).
\end{equation}
To explain this embedding, note that
in general, a  Lie algebra of contact vector fields
can be realized as a
subalgebra of Poisson algebra [6]. 
In particular, the Lie algebra $Vect(S^1)$
of smooth vector fields on the circle can be embedded
into the Poisson algebra of functions on 
the cylinder $\dot{T}^*S^1 = T^*S^1\setminus S^1$
with the removed zero section.
One can  introduce 
the Darboux coordinates
$(q, p) = (t, \xi)$ on this manifold.
The symbols of differential operators are functions on  $\dot{T}^*S^1$
which are formal Laurent series in $p$ with coefficients
periodic in $q$. Correspondingly, 
they define Hamiltonian vector fields on $\dot{T}^*S^1$ [5].
Then  $Vect(S^1)$ is realized as 
the subalgebra of the Lie algebra of
Hamiltonian vector fields, consisting of the fields  with Hamiltonians 
which are homogeneous of degree 1.
This condition holds in general, if one considers the
{\it symplectification} of a contact manifold [6], and
it can be generalized to the supercase.

Let  $P(N) = \oplus_{j\in\mathbb Z}P^j(N)$ be the 
$\mathbb Z$-grading of the associative superalgebra $P(N)$
defined by
$\hbox{deg }\xi = 
\hbox{deg } \bar{\theta}_i = 1,  
\hbox{deg } t =
\hbox{deg } {\theta}_i = 0 \hbox{ for }i = 1, \ldots, N.$
With respect to the Poisson bracket,
$\lbrace P^j(N), P^k(N)\rbrace \subset P^{j+k-1}(N).$
Then $P^1(N)$ is a subalgebra of $P(N)$, and it is isomorphic to $K(2N)$ [8].

{\bf 3.-} 
A natural question is whether there exists an embedding
of $K(2N)$ into $P_h(N)$.
If $N = 1$, then  $P^1(1) = W(1)\cong K(2)$. Thus, clearly, $K(2)\subset P_h(1)$.
If $N = 2$, then  $K'(4)\subset P(2)$
is defined from the exact sequence 
\begin{equation}
0\rightarrow K'(4)\rightarrow P^1(2)\rightarrow \mathbb C 
{t^{-1} \xi^{-1}\theta_1\theta_2\bar{\theta}_1\bar{\theta}_2}\rightarrow 0.
\end{equation}
The 2-cocycles
on $K'(4)$ in this realization are defined as follows. Let
\begin{eqnarray}
c(t^n\xi, t^k\xi^{-1}{\theta}_1{\theta}_2
\bar{\theta}_1\bar{\theta}_2) &=& A\delta_{n+k, 0}, n\not= 1,\nonumber\\
c(t^n\xi{\theta}_i, t^k\xi^{-1}{\theta}_j
\bar{\theta}_1\bar{\theta}_2) &=& A(-1)^j\delta_{n+k, 0}, i\not= j,\\
c(t^n\xi{\theta}_1{\theta}_2, t^k\xi^{-1}
\bar{\theta}_1\bar{\theta}_2) &=& A\delta_{n+k, 0},\nonumber\
\end{eqnarray}
\begin{eqnarray}
c(t^n\bar{\theta}_i, t^k{\theta}_1{\theta}_2\bar{\theta}_j) &=&
B(-1)^j\delta_{n+k, 0}, i\not= j,\nonumber\\
c(t^n{\theta}_1\bar{\theta}_i, t^k{\theta}_2\bar{\theta}_j) &=&
B(-1)^i\delta_{n+k, 0}, i\not= j,\nonumber\
\end{eqnarray}
\begin{eqnarray}
c(t^n\xi^{-1}{\theta}_1{\theta}_2\bar{\theta}_1\bar{\theta}_2,
t^k{\theta}_i\bar{\theta}_i) &=&  C\delta_{n+k+1, 0}, n\not= -1,\nonumber\\
c(t^n\xi^{-1}{\theta}_i\bar{\theta}_1\bar{\theta}_2,
t^k{\theta}_1{\theta}_2\bar{\theta}_i) &=&  -C\delta_{n+k+1, 0},\nonumber\\
c(t^n\xi^{-1}\theta_1\theta_2\bar{\theta}_1\bar{\theta}_2,
t^k\xi^{-1}\theta_1\theta_2\bar{\theta}_1\bar{\theta}_2) &=&
-\frac {2C}{n+1}\delta_{n+k+2, 0}, n \not= -1,\nonumber\
\end{eqnarray}
where $A, B, C \in \mathbb Z$.
Then three linearly independent 2-cocycles $c_1$, $c_2$, and $c_3$
on $K'(4)$ are given by 
\begin{eqnarray}
&c_1:& A = 1, B = 0, C = 0,\nonumber\\
&c_2:& A = n, B = n, C = 0,\\
&c_3:& A = n, B = 0, C = 1.\nonumber\
\end{eqnarray}
Note that these cocycles are represented by linear functions of the indexes.
In [1] and  [3] another formulas for cocycles are given, which involve
the usual ``cubic term'' of the Virasoro cocycle.
Note that the exterior derivations of $K'(4)$ are
$Der_{ext}K'(4) = \mathbb C D$, where $D$ is the derivation
defined by (13):
$D(x) = \lbrace t^{-1} \xi^{-1}\theta_1\theta_2\bar{\theta}_1\bar{\theta}_2,
x\rbrace$ for $x\in K'(4)$.
There is a natural action of $Der_{ext}K'(4)$ on
the cohomology space $H^2(K'(4), \mathbb C)$. Under this action 
$D(c_1) = D(c_2) = 0, D(c_3) = -2c_1$.
Correspondingly, the exterior automorphism $Exp D$ of $K'(4)$ 
associates the cocycles $c_3$ and $c_3 - 2c_1$.
Thus, up to automorphisms, there are two independent 2-cocycles 
on $K'(4)$: $c_1$ and $c_2$ (cf. [7]).

The embedding $K'(4)\subset P(2)$
can be obtained as follows. Consider the one-parameter
family  $S(2, \alpha)$ ($\alpha\in\mathbb C$) of deformations of
the  Lie superalgebra of divergence-free 
derivations of $\Lambda (1, 2)$ [1].
The exterior derivations of 
$S'(2, \alpha)$ form an $\mathfrak s\mathfrak l(2)$ if $\alpha\in\mathbb Z$.
The exterior derivations of $S'(2, \alpha)$  for all
$\alpha\in\mathbb Z$ generate a subalgebra of $P(2)$
isomorphic to the loop algebra
$\tilde {\mathfrak  s\mathfrak  l}(2)$ 
($\mathfrak s\mathfrak l(2)$ corresponds to $\alpha = 1$).
The family
$S'(2, \alpha)$ for all $\alpha \in \mathbb Z$
and $\tilde{\mathfrak s\mathfrak l}(2)$ generate a Lie superalgebra 
isomorphic to $K'(4)$.
The similar construction for each $h\in \rbrack 0, 1]$ gives
an embedding of a nontrivial central extension of $K'(4)$
\begin{eqnarray}
\hat{K}'(4) \subset P_h(2)
\end{eqnarray}
such that the central element is $h\in P_h(2)$, and
$\hbox{lim}_{h\rightarrow 0}\hat{K}'(4) =  K'(4)\subset P(2)$ [8].
The superalgebra $K'(4)\subset P(2)$ is spanned 
by $W(2)$ together with four fields $F^i_n$, where
$i = 0, 1, 2, 3$, and $n \in \mathbb Z$:
\begin{eqnarray}
F^0_n &=& t^n\xi^{-1}\bar{\theta}_1\bar{\theta}_2,\nonumber\\
F^i_n &=&t^n\xi^{-1}{\theta}_i\bar{\theta}_1\bar{\theta}_2, \quad i = 1, 2,\\
F^3_n &=& t^n\xi^{-1}{\theta}_1{\theta}_2
\bar{\theta}_1\bar{\theta}_2,\quad n\not= -1.\nonumber
\end{eqnarray}
The superalgebra $\hat{K}'(4)\subset P_h(2)$ is spanned 
by $W(2)$ together with four fields $F^i_{n,h}$:
\begin{eqnarray}
F^0_{n,h} &=& (\xi^{-1}\circ_h t^n)\partial_1\partial_2,\nonumber\\
F^i_{n,h} &=& (\xi^{-1}\circ_h t^n)\partial_1\partial_2{\theta}_i,\quad i = 1, 2,\\
F^3_{n,h} &=& (\xi^{-1}\circ_h t^n)\partial_1\partial_2{\theta}_1{\theta}_2
+ \frac {h} {n+1}t^{n+1}, \quad n\not= -1,\nonumber
\end{eqnarray}
and the central element $h$.
The corresponding 2-cocycle on $K'(4)$ is $c_1$. Note that
the cocycles $c_1$ and $c_2$ are equal, respectively,
to the restrictions of the cocycles $c_t$ and $c_{\xi}$ on $P_{h=1}(2)$.

The embedding $(16)$ in the case where $h = 1$
allows to obtain a one-parameter family of irreducible
representations of $\hat{K}'(4)$
in the superspace spanned by 2 even fields and 2 odd fields
where the value of the central charge is equal to one [8].
To describe this representation,  consider
the superspace $V^{\mu}$ spanned by the functions $t^{\mu}g$,
where $g\in \Lambda (1, 2)$, and $\mu \in \mathbb R \setminus\mathbb Z$.
Let $\lbrace v_m^i\rbrace$, where ${m\in \mathbb Z}$ and $i = 0, 1, 2, 3$,
be the following basis in $V^{\mu}$:
\begin{equation}
v_m^0 = \frac {1} {m + \mu}t^{m + \mu}, \quad
v_m^1 = t^{m + \mu}\theta_1,\quad
v_m^2 = t^{m + \mu}\theta_2, \quad
v_m^{3} = t^{m + \mu}\theta_1\theta_2. 
\end{equation}
Every $D \in W(2)$ is  a derivation of $V^{\mu}$.
To define an action of $F^i_{n,1}$ on  $V^{\mu}$, one can interpret
$\xi^{-1}$ as the anti-derivative 
on the space of functions $t^\mu\mathbb C [t, t^{-1}]$.  Then 
\begin{eqnarray}
F^0_{n,1}(v_m^3) &=& - v_{m+n+1}^0, \nonumber\\
F^1_{n,1}(v_m^2) &=& - v_{m+n+1}^0,\quad 
F^2_{n,1}(v_m^1) =  v_{m+n+1}^0, \\
F^3_{n,1}(v_m^i) &=& \frac {1} {n+1} v_{m+n+1}^i, \quad i = 0, 1, 2, 3; 
\quad n\not= -1.\nonumber
\end{eqnarray}
The central element in $\hat{K}'(4)$ is $1$, and it acts by the identity 
operator. 
In this way one obtains a family of representations of 
$\hat{K}'(4)$
in the superspace spanned by four fields $v_m^i$, where $i = 0, 1, 2, 3$,
and $\mu$ appears as an
arbitrary complex parameter [8].

No analogous embedding of  $K(2N)$ into $P_h(N)$
exists if $N\geq 3$.

\vskip 0.2in

BIBLIOGRAPHY.

\vskip 0.1in

[1] V. G. Kac and J. W. van de Leur,
{\it On classification of superconformal algebras}, 
in {\it Strings-88}, edited by S. J. Gates  et al.,
World Scientific, Singapore, 1989,  77-106.

[2] V. G. Kac,
{\it Superconformal algebras and transitive group actions on quadrics},
Commun. Math. Phys. {\bf 186}, (1997) 233-252.

[3] P. Grozman, D. Leites, and I. Shchepochkina,
{\it Lie superalgebras of string theories},
hep-th/9702120.

[4] M. Ademollo, L. Brink, A. D'Adda  et al.,
{\it Dual strings with $U(1)$ colour symmetry}, 
Nucl. Phys. B {\bf 111}, (1976) 77-110.

[5] B. Khesin, V. Lyubashenko, and C. Roger,
{\it Extensions and contractions of the Lie algebra of
q-pseudodifferential symbols on the circle},
J. Funct. Anal. {\bf 143}, (1997) 55-97.

[6]  V. I. Arnold, 
{\it Mathematical Methods of Classical Mechanics},
Springer-Verlag, New York, 1989.

[7] K. Schoutens,
{\it $O(N)$-extended
superconformal field theory in superspace},
Nucl. Phys. B {\bf 295}, (1988) 634-652.

[8] E. Poletaeva,
{\it A spinor-like representation of the contact
superconformal algebra  $K'(4)$}, 
J. Math. Phys. {\bf 42}, (2001) 526-540; hep-th/0011100
and references therein.

\vskip 0.1in

Elena Poletaeva

Centre for Mathematical Sciences

Mathematics, Lund University

Box 118, S-221 00 Lund, Sweden

elena$@$maths.lth.se

\end{document}